\documentclass[twocolumn,showpacs,amsmath,amssymb,10pt]{revtex4}
\usepackage{amsfonts}
\usepackage{bm}

\newcommand{\ot}{{\,\otimes\,}}
\newcommand{{\Cd}}{{\mathbb{C}^d}}

\def\oper{{\mathchoice{\rm 1\mskip-4mu l}{\rm 1\mskip-4mu l}%
{\rm 1\mskip-4.5mu l}{\rm 1\mskip-5mu l}}}
\def\<{\langle}
\def\>{\rangle}
\newtheorem{theorem}{Theorem}

\newtheorem{lemma}{Lemma}

\begin{document}
\title{\textbf{Spectral conditions for entanglement witnesses vs. bound entanglement}} \author{Dariusz
Chru\'sci\'nski, Andrzej Kossakowski and Gniewomir
Sarbicki\thanks{email: darch@phys.uni.torun.pl} }
\affiliation{Institute of Physics, Nicolaus Copernicus University,\\
Grudzi\c{a}dzka 5/7, 87--100 Toru\'n, Poland}

\begin{abstract}

It is shown that  entanglement witnesses constructed via the family
of spectral conditions are decomposable, i.e. cannot be used to
detect bound entanglement. It supports several observations that
bound entanglement reveals highly non-spectral features.

\end{abstract}
\pacs{03.65.Ud, 03.67.-a}

\maketitle

\section{Introduction}

One of the most important problems of quantum information theory
\cite{QIT,HHHH} is the characterization of mixed states of composed
quantum systems. In particular it is of primary importance to test
whether a given quantum state exhibits quantum correlation, i.e.
whether it is separable or entangled. For low dimensional systems
there exists simple necessary and sufficient condition for
separability. The celebrated Peres-Horodecki criterium \cite{PPT}
states that a state of a bipartite system living in $\mathbb{C}^2
\ot \mathbb{C}^2$ or $\mathbb{C}^2 \ot \mathbb{C}^3$ is separable
iff its partial transpose is positive. Unfortunately, for
higher-dimensional systems there is no single {\it universal}
separability condition. Apart from PPT criterion there are several
separability criteria available in the literature (see \cite{HHHH}
and \cite{Geza} for the review). However, each of them defines only
a necessary condition.

The power and simplicity of Peres-Horodecki criterion comes from the
fact that it is based on the spectral property: to check for PPT one
simply checks the spectrum of $\rho^\Gamma = (\oper \ot T)\rho$.
Another simple spectral separability test is known as the reduction
criterion \cite{RED}
\begin{equation}\label{R}
\mathbb{I}_A \ot \rho_B \geq \rho\ , \ \ {\rm and} \ \ \rho_A \ot
\mathbb{I}_B \geq \rho\ ,
\end{equation}
where $\rho_A = {\rm Tr}_B\rho$ ($\rho_B = {\rm Tr}_A\rho$) is the
reduced density operator. However, reduction criterion is weaker
that Peres-Horodecki one, i.e. any PPT state does satisfy (\ref{R})
as well.

 Actually, there exist other criteria which
are based on spectral properties. For example it turns out that
separable states satisfy so called entropic inequalities
\begin{equation}\label{S}
S(\rho) - S(\rho_A) \geq 0\ , \ \ {\rm and}\ \  S(\rho) - S(\rho_B)
\geq 0\ ,
\end{equation}
where $S$ denotes the von Neumann entropy.   This means that in the
case of separable states the whole system is more disordered than
its subsystems. Actually, these inequalities may be generalized
\cite{E1,E2,E3} for R\'enyi entropy  (or equivalently Tsallis
entropy). Another spectral tool was proposed by Nielsen and Kempe
\cite{Kempe} and it is based on the majorization criterion
\begin{equation}\label{M}
    \lambda(\rho_A) \succ \lambda(\rho)\ , \ \ {\rm and} \ \
    \lambda(\rho_B) \succ \lambda(\rho)\ ,
\end{equation}
where $\lambda(\rho)$ and $\lambda(\rho_{A(B)})$  denote vectors
consisting of eigenvalues of $\rho$ and $\rho_{A(B)}$, respectively.
Recall, that if $x=(x_1,\ldots,x_n)$ and $y=(y_1,\ldots,y_n)$ are
two stochastic vectors, then $x \prec y$ if
\begin{equation}\label{}
    \sum_{i=1}^k x_i^\downarrow \leq  \sum_{i=1}^k y_i^\downarrow \
    ,\ \ \ \ k=1,\ldots,n-1\ ,
\end{equation}
where $x^\downarrow$  ($1 \leq i \leq n$) are components of vector
$x$ rearranged in decreasing order ($x^\downarrow_1 \geq \ldots \geq
x_n^\downarrow$) and similarly for $y^\downarrow_i$. Actually,
majorization can  be shown \cite{Bhatia} to be a more stringent
notion of disorder than entropy in the sense that if $x \prec y$,
then it follows that $H(x) \geq H(y)$, where $H(x)$ stands for the
Shanon entropy of the stochastic vector $x$.

Interestingly, both criteria, i.e. entropic inequalities (\ref{S})
and majorization relations (\ref{M}) follow from the reduction
criterion (\ref{R}) \cite{E3,Hiroshima}. It means that they cannot
be used to detect bound entanglement. In particular, since PPT
criterion $\rho^\Gamma \geq 0$ implies (\ref{R}), the above spectral
tests are useless in searching for PPT entangled states.

The most general approach to characterize quantum entanglement uses
a notion of an entanglement witness (EW) \cite{HHH,Terhal1}. A
Hermitian operator $W$ defined on a tensor product
$\mathcal{H}=\mathcal{H}_A \ot \mathcal{H}_B$ is called  an EW iff
1) $\mbox{Tr}(W\sigma_{\rm sep})\geq 0$ for all separable states
$\sigma_{\rm sep}$, and 2) there exists an entangled state $\rho$
such that $\mbox{Tr}(W\rho)<0$ (one says that $\rho$ is detected by
$W$). It turns out that a state is entangled if and only if it is
detected by some EW \cite{HHH}. There was a considerable effort in
constructing and analyzing the structure of EWs
\cite{EW1,Bruss,Toth,nasze,Gniewko} (see also \cite{HHHH} for the
review). However, the general construction of these objects is not
known.

 In the recent
paper \cite{CMP} we proposed a new class of entanglement witnesses.
Their construction is based on the family of spectral conditions.
Therefore, they do belong to the family of spectral separability
tests. This class recovers many well known examples of entanglement
witnesses. In the present paper we show that similarly to other
spectral tests our new class of witnesses cannot be used to detect
PPT entangled states. It means that these witnesses are
decomposable.

The paper is organized as follows: in the next Section we recall the
construction of entanglement witnesses from \cite{CMP}.
Section~\ref{EX} presents several examples from the literature which
do fit our class. Section~\ref{MAIN} contains our main result --
proof of decomposability. Final conclusions are collected in the
last Section.

\section{Construction of the spectral class}

Any entanglement witness $W$ can be represented as a difference $W =
W_+ - W_-$, where both $W_+$ and $W_-$ are semi-positive operators
in $\mathcal{B}(\mathcal{H}_A \ot \mathcal{H}_B)$. However, there is
no general method to recognize that $W$ defined by $W_+ - W_-$ is
indeed an EW. One particular method based on spectral properties of
$W$ was presented in \cite{CMP}. Let $\psi_\alpha$ ($\alpha
=1,\ldots,D=d_Ad_B$) be an orthonormal basis in $\mathcal{H}_A \ot
\mathcal{H}_B$ and denote by $P_\alpha$ the corresponding projector
$P_\alpha = |\psi_\alpha\>\<\psi_\alpha|$. It leads therefore to the
following spectral resolution of identity
\begin{equation}\label{}
\mathbb{I}_A \ot \mathbb{I}_B = \sum_{\alpha=1}^D P_\alpha\ .
\end{equation}
Now, take $D$  semi-positive numbers $\lambda_\alpha \geq 0$ such
that $\lambda_\alpha$ is strictly positive for $\alpha > L$, and
define
\begin{equation}\label{}
    W_- = \sum_{\alpha=1}^L \lambda_\alpha P_\alpha\ , \ \ \ \
W_+ = \sum_{\alpha=L+1}^D \lambda_\alpha P_\alpha\ ,
\end{equation}
where $L$ is an arbitrary integer $0<L<D$. This construction
guarantees that $W_+$ is strictly positive and all zero modes and
strictly negative eigenvalues of $W$ are encoded into $W_-$.
Consider normalized vector $\psi \in \mathcal{H}_A \ot
\mathcal{H}_B$ and let
\[ s_1(\psi) \geq \ldots \geq s_d(\psi)  \ ,\]
denote its Schmidt coefficients $(d=\min\{d_A,d_B\})$. For any $1
\leq k \leq d$ one defines  $k$-norm of $\psi$ by the following
formula \cite{KF}
\begin{equation}\label{}
    || \psi ||^2_k = \sum_{j=1}^k s^2_j(\psi)\ .
\end{equation}
It is clear that
\begin{equation}\label{}
    ||\psi ||_1 \leq ||\psi||_2 \leq \ldots \leq ||\psi ||_d \ .
\end{equation}
Note that $||\psi||_1$ gives the maximal Schmidt coefficient of
$\psi$, whereas due to the normalization, $||\psi||^2_d =
\<\psi|\psi\> =1$. In particular, if $\psi$ is maximally entangled
then
\begin{equation}\label{}
    ||\psi||^2_k = \frac{k}{d} \ .
\end{equation}
Equivalently one may define $k$-norm of $\psi$ by
\begin{equation}\label{}
    ||\psi||^2_k = \max_\phi |\<\psi|\phi\>|^2\ ,
\end{equation}
where the maximum runs over all normalized vectors $\phi$ such that
${\rm SR}(\psi) \leq k$ (such $\phi$ is usually called
$k$-separable). Recall that a Schmidt rank of $\psi$ -- ${\rm
SR}(\psi)$ -- is the number of non-vanishing Schmidt coefficients of
$\psi$. One calls entanglement witness $W$ a $k$-EW if
$\<\psi|W|\psi\> \geq 0 $ for all $\psi$ such that ${\rm
SR}(\psi)\leq k$.
 The main  result of
\cite{CMP} consists in the following
\begin{theorem} Let $\sum_{\alpha=1}^L ||\psi_\alpha||^2_{k} < 1$.
If the following spectral conditions are satisfied
\begin{equation} \label{T1}
     \lambda_\alpha \geq \mu_k\ , \ \ \
    \alpha=L+1,\ldots,D\ ,
\end{equation}
where
\begin{equation}  \label{mu-l}
    \mu_\ell := \frac{\sum_{\alpha=1}^L \lambda_\alpha
    ||\psi_\alpha||^2_\ell}{1-\sum_{\alpha=1}^L
    ||\psi_\alpha||^2_\ell}\ ,
\end{equation}
then $W$ is an $k$-EW. If moreover $\sum_{\alpha=1}^L
||\psi_\alpha||^2_{k+1} < 1$ and
\begin{equation}\label{T2}
    \mu_{k+1} > \lambda_\alpha\ , \ \ \
    \alpha=L+1,\ldots,D\ ,
\end{equation}
then $W$ being $k$-EW is not $(k+1)$-EW.
\end{theorem}

\section{Examples}  \label{EX}

Surprisingly this simple construction recovers many well know
examples of EWs.

{\it Example 1}. Flip operator in $d_A=d_B=2$:
\begin{equation}\label{Flip}
    W = \left( \begin{array}{c c|c c} 1 & \cdot & \cdot & \cdot \\ \cdot & \cdot & 1 & \cdot \\
    \hline \cdot & 1 & \cdot  & \cdot \\ \cdot & \cdot & \cdot & 1 \end{array}
    \right)\ ,
\end{equation}
where dots represent zeros. Its spectral decomposition has the
following form: $W_-=\lambda_1 P_1$
\begin{equation*}\label{}
    \lambda_1=\lambda_2=\lambda_3=\lambda_4=1\ ,
\end{equation*}
and
\begin{eqnarray*}
  \psi_1 &=& \frac{1}{\sqrt{2}} (|12\> - |21\> ) \ , \\
  \psi_2 &=& \frac{1}{\sqrt{2}} (|12\> + |21\> ) \ , \\
  \psi_3 &=& |11\> \ , \ \ \psi_4 \ = \ |22\>\ .
\end{eqnarray*}
One finds $\mu_1=1$ and hence condition (\ref{T1}) is trivially
satisfied $\lambda_\alpha \geq \mu_1$ for $\alpha =2,3,4$. We stress
that our construction does not recover flip operator in $d>2$. It
has $d(d-1)/2$ negative eigenvalues. Our construction leads to at
most $d-1$ negative eigenvalues.

{\it Example 2:} Entanglement witness corresponding to the reduction
map:
\[ \lambda_1 = d-1, \ \ \lambda_2 = \ldots = \lambda_D = 1  \ , \]
and
\begin{equation}\label{}
    W_- = P^+_d\ , \ \ \ W_+ = \mathbb{I}_d \ot \mathbb{I}_d -
    P^+_d\ ,
\end{equation}
where $P^+_d$ denotes maximally entangled state in $\mathbb{C}^d \ot
\mathbb{C}^d$. Again, one finds $\mu_1=1$ and hence condition
(\ref{T1}) is trivially satisfied $\lambda_\alpha \geq \mu_1$ for
$\alpha =2,\ldots,D=d^2$. Now, since $\psi_1$ corresponds to the
maximally entangled state one has $1 - ||\psi_1||^2_2 = (d-2)/d<1$.
Hence, condition (\ref{T2})
\begin{equation}\label{}
    \mu_2 = 2 \frac{d-2}{d-1} > \lambda_\alpha \ , \ \ \alpha
    =2,\ldots,D\ ,
\end{equation}
implies that $W$ is not a 2-EW.

{\it Example 3:} A family of $k$-EW in $\mathbb{C}^d \ot
\mathbb{C}^d$ defined by \cite{SN}
\[ \lambda_1 = pd-1, \ \ \lambda_2 = \ldots = \lambda_D = 1  \ , \]
with $p\geq 1$, and
\begin{equation}\label{}
    W_- = P^+_d\ , \ \ \ W_+ = \mathbb{I}_d \ot \mathbb{I}_d -
    P^+_d\ .
\end{equation}
Clearly, for $p=1$ it reproduces the reduction EW. Now, conditions
(\ref{T1}) and (\ref{T2}) imply that if
\begin{equation}\label{}
    \frac{1}{k+1} < p \leq \frac{1}{k}\ ,
\end{equation}
then $W$ is $k$- but not $(k+1)$-EW.

{\it Example 4:} A family of EWs in $\mathbb{C}^3 \ot \mathbb{C}^3$
defined by \cite{Cho-Kye}
\begin{equation}\label{}
  W[a,b,c]\, =\,  \left( \begin{array}{ccc|ccc|ccc}
    a & \cdot & \cdot & \cdot & -1 & \cdot & \cdot & \cdot & -1 \\
    \cdot& b & \cdot & \cdot & \cdot& \cdot & \cdot & \cdot & \cdot  \\
    \cdot& \cdot & c & \cdot & \cdot & \cdot & \cdot & \cdot &\cdot   \\ \hline
    \cdot & \cdot & \cdot & c & \cdot & \cdot & \cdot & \cdot & \cdot \\
    -1 & \cdot & \cdot & \cdot & a & \cdot & \cdot & \cdot & -1  \\
    \cdot& \cdot & \cdot & \cdot & \cdot & b & \cdot & \cdot & \cdot  \\ \hline
    \cdot & \cdot & \cdot & \cdot& \cdot & \cdot & b & \cdot & \cdot \\
    \cdot& \cdot & \cdot & \cdot & \cdot& \cdot & \cdot & c & \cdot  \\
    -1 & \cdot& \cdot & \cdot & -1 & \cdot& \cdot & \cdot & a
     \end{array} \right)\ ,
\end{equation}
with $a,b,c\geq 0$. Necessary and sufficient conditions for
$W[a,b,c]$ to be an EW are
\begin{enumerate}
\item $0 \leq a < 2\ $,
\item $ a+b+c \geq 2\ $,
\item if $a \leq 1\ $, then $ \ bc \geq (1-a)^2$.
\end{enumerate}
 A family $W[a,b,c]$ generalizes celebrated Choi
indecomposable witness corresponding to $a=b=1$ and $c=0$. Now,
spectral properties of $W[a,b,c]=W_+ - W_-$ read as follows: $W_-=
\lambda_1 P^+_3$ and
\[  \lambda_1 = 2-a\ ,\  \ \lambda_2=\lambda_3 = a+1 \ ,\]
\[  \lambda_4=\lambda_5=\lambda_6=b \ , \ \  \lambda_7=\lambda_8=\lambda_9=c\ .\]
One finds $\mu_1 = (2-a)/2 $ and hence condition (\ref{T1}) implies
\begin{equation}\label{}
    a\geq 0\ , \ \ b,c \geq \frac{2-a}{2} \ .
\end{equation}
It gives therefore
\begin{equation}\label{}
    a+b+c \geq  2\ ,
\end{equation}
and one easily shows that the  conditions 3 is also satisfied.
Summarizing: $W[a,b,c]$ belongs to our spectral class if and only if
\begin{enumerate}
\item $0 \leq a < 2\ $,
\item $ b,c \geq (2-a)/2\ $ .
\end{enumerate}
Note that the Choi witness $W[1,1,0]$ does not belong to this class.
It was shown \cite{Cho-Kye} that $W[a,b,c]$ is decomposable if and
only if $a\geq 0$ and
\begin{equation}\label{DEC}
     bc \geq \frac{(2-a)^2}{4}\ .
\end{equation}
Hence $W[a,b,c]$ from our spectral class is always decomposable. In
particular $W[0,1,1]$ reproduces the EW corresponding to the
reduction map in $d=3$.  Note, that there are entanglement witnesses
$W[a,b,c]$ which are decomposable, i.e. satisfy (\ref{DEC}), but do
not belong to or spectral class. Similarly one can check when
$W[a,b,c]$ defines 2-EW. One finds $\mu_2 = 2(2-a)$ and hence
condition (\ref{T1}) implies
\begin{enumerate}
\item $1 \leq a < 2\ $,
\item $ b,c \geq 2(2-a)\ $ .
\end{enumerate}
Clearly, any 2-EW from our class is necessarily decomposable. It was
shown \cite{Cho-Kye} that all 2-EW from the class $W[a,b,c]$ are
decomposable.

Interestingly all examples 1--4 show one characteristic feature --
entanglement witnesses satisfying spectral conditions (\ref{T1}) are
decomposable. In the next Section we show that it is not an
accident.


\section{Decomposability of the spectral class}  \label{MAIN}

Indeed,  we show that if entanglement witness $W$ does satisfy
(\ref{T1}) with $k=1$, then it is necessarily decomposable. It means
that if $\rho$ is PPT, then it cannot be detected by $W$:
\begin{equation}\label{XX}
    \rho^\Gamma \geq 0 \ \ \Longrightarrow\ \ {\rm Tr}(\rho W) \geq
    0 \ .
\end{equation}
To prove it note that
\begin{equation}\label{W-AB}
    W = A + B \ ,
\end{equation}
where
\begin{equation}\label{}
A = \sum_{\alpha=L+1}^D
    (\lambda_\alpha - \mu_1) P_\alpha\ ,
\end{equation}
and
\begin{equation}\label{}
    B = \mu_1 \mathbb{I}_A \ot \mathbb{I}_B  - \sum_{\alpha=1}^L
    (\lambda_\alpha + \mu_1) P_\alpha\ .
\end{equation}
Now, since $\lambda_\alpha \geq \mu_1$, for $\alpha =L+1,\ldots,D$,
it is clear that $A \geq 0$. The partial transposition of $B$ reads
as follows
\begin{equation}\label{}
 B^\Gamma = \mu_1 \mathbb{I}_A \ot \mathbb{I}_B  - \sum_{\alpha=1}^L
    (\lambda_\alpha + \mu_1) P_\alpha^\Gamma\ .
\end{equation}
Let us recall that the spectrum of the partial transposition of
rank-1 projector $|\psi\>\<\psi|$ is well know: the nonvanishing
eigenvalues of $|\psi\>\<\psi|^\Gamma$ are given by
$s_\alpha^2(\psi)$ and $\pm s_\alpha(\psi) s_\beta(\psi) $, where
$s_1(\psi) \geq \ldots \geq s_d(\psi)$ are Schmidt coefficients of
$\psi$. Therefore, the smallest eigenvalue of $B^\Gamma$ (call it
$b_{\rm min}$) satisfies
\begin{equation}\label{}
    b_{\rm min} \geq \mu_1 - \sum_{\alpha=1}^L
    (\lambda_\alpha + \mu_1) ||\psi_\alpha||_1^2\ ,
\end{equation}
and using the definition of $\mu_1$ (cf. Eq. (\ref{mu-l})) one gets
\begin{equation}\label{}
    b_{\rm min} \geq 0 \ ,
\end{equation}
which implies $B^\Gamma \geq 0$. Hence, due to the formula
(\ref{W-AB}) the entanglement witness $W$ is decomposable.

Interestingly, saturating the bound (\ref{T1}), i.e. taking
\begin{equation}\label{SAT}
    \lambda_\alpha = \mu_1\ , \ \ \  \alpha=L+1,\ldots, D\ ,
\end{equation}
one has $A=0$ and hence $W = Q^\Gamma$ with $Q = B^\Gamma \geq 0$
which shows that the corresponding positive map $\Lambda :
\mathcal{B}(\mathcal{H}_A) \rightarrow \mathcal{B}(\mathcal{H}_B)$
defined by
\begin{equation}\label{}
    \Lambda(X) = {\rm Tr}_A (W \cdot  X^T \ot \mathbb{I}_B) \ ,
\end{equation}
is completely co-positive, i.e. $\Lambda \circ T$ is completely
positive. Note that
\begin{equation}\label{Lambda}
    \Lambda(X) = \mu_1 \mathbb{I}_B {\rm Tr}\, X - \sum_{\alpha=1}^L
    (\mu_1+\lambda_\alpha) F_\alpha X F_\alpha^\dagger \ ,
\end{equation}
where $F_\alpha$ is a linear operator $F_\alpha : \mathcal{H}_A
\rightarrow \mathcal{H}_B$ defined by
\begin{equation}\label{}
    \psi_\alpha = \sum_{i=1}^{d_A} e_i \ot F_\alpha e_i \ ,
\end{equation}
and $\{e_1,\ldots,e_{d_A}\}$ denotes an orthonormal basis in
$\mathcal{H}_A$. In particular, if $L=1$, i.e. there is only one
negative eigenvalue, then formula (\ref{Lambda}) (up to trivial
rescaling) gives
\begin{equation}\label{Lambda-1}
    \Lambda(X) = \kappa\, \mathbb{I}_B {\rm Tr}\, X -  F_1 X  F_1^\dagger \ ,
\end{equation}
with
\begin{equation}\label{}
    \kappa = \frac{\mu_1}{\mu_1 + \lambda_1} = ||\psi_1||_1^2\ .
\end{equation}
It reproduces a positive map (or equivalently an EW $W = \kappa\,
\mathbb{I}_A \ot \mathbb{I}_B - P_1$) which is known to be
completely co-positive \cite{Geza,Marco,Gniewko}. If $d_A=d_B=d$ and
$\psi_1$ is maximally entangled, that is, $F_1 = U/\sqrt{d}$ for
some unitary $U \in U(d)$, then one finds for $\kappa = 1/d$ and the
map (\ref{Lambda-1}) is unitary equivalent to the reduction map
$\Lambda(X) = U R(X) U^\dagger$, where $R(X) = \mathbb{I}_d {\rm
Tr}X - X$.

Finally, let us observe that EWs presented in Examples 1-3 are not
only decomposable but completely co-positive, i.e. $W^\Gamma \geq
0$. Moreover, the flip operator (\ref{Flip}) and the EW
corresponding to the reduction map do satisfy (\ref{SAT}). EW from
Example 4 fitting our spectral class is in general only decomposable
but $W[a,b,c]^\Gamma \ngeqslant 0$. Its partial transposition
becomes positive if in addition to $ b,c \geq (2-a)/2\ $ it
satisfies $bc\geq 1$. Note, that condition (\ref{SAT}) implies in
this case
\[  b=c=a+1 = \frac{2-a}{2}  \ , \]
which leads to $a=0$ and $b=c=1$. This case, however, corresponds to
the standard reduction map in $\mathbb{C}^3$.

\section{Conclusions}

We have shown that the spectral class of entanglement witnesses
constructed recently in \cite{CMP} contains only decomposable EWs,
that is, it cannot be used to detect PPT entangled state. This
observation supports other results like entropic inequalities
(\ref{S}) and majorization relations (\ref{M}) which are also
defined via spectral conditions and  turned out to be unable to
detect bound entanglement. We conjecture that ``spectral tools" are
inappropriate in searching for bound entanglement which shows highly
non-spectral features.

\section*{Acknowledgements}

We thank Marco Piani for valuable comments and pointing out
references \cite{Geza} and \cite{Marco}. This work was partially
supported by the Polish Ministry of Science and Higher Education
Grant No 3004/B/H03/2007/33 and by the Polish Research Network  {\it
Laboratory of Physical Foundations of Information Processing}.


\begin{thebibliography}{1}
\bibitem{KF} Actually, this family of $k$-norms is equaivalent to the well
known Ky-Fan $k$-norms in the space of bounded linear operators from
$\mathcal{H}_A$ to $\mathcal{H}_B$ (see \cite{CMP} for details).
\end{thebibliography}
\end{document}